\documentclass[aps, prr, twocolumn, superscriptaddress, amsmath, tightenlines, longbibliography]{revtex4-1}

\usepackage{dcolumn}
\usepackage{graphicx}
\usepackage{mathrsfs}
\usepackage{subfigure}
\usepackage{booktabs}
\usepackage{amsmath}
\usepackage{physics}
\usepackage{dsfont}
\usepackage{amstext}
\usepackage{amssymb}
\usepackage{amsbsy}
\usepackage{bbm}
\usepackage{amsthm}
\usepackage{graphicx}
\usepackage{color}
\usepackage[colorlinks,citecolor=blue]{hyperref}

\usepackage{url}
\usepackage[colorlinks]{hyperref}
\hypersetup{%
	plainpages=true,
	breaklinks=true,       
	hypertexnames=false,  
	pageanchor=true,
	colorlinks=true,
	linkcolor={blue},
	citecolor={red},
	urlcolor={blue},
	anchorcolor={black}
}

 \makeatletter

\newcommand{\Rmnum}[1]{\expandafter\@slowromancap\romannumeral #1@}
\makeatother

\begin{document}
	
\title{Arbitrary Control of Non-Hermitian Skin Modes via Disorder and An Electric Field}
\author{Zhao-Fan Cai}
\thanks{These authors contributed equally}
\affiliation{School of Physics and Optoelectronics, South China University of Technology,  Guangzhou 510640, China}
\author{Yang Li}
\thanks{These authors contributed equally}
\affiliation{School of Physics and Optoelectronics, South China University of Technology,  Guangzhou 510640, China}
\author{Yu-Ran Zhang}
\affiliation{School of Physics and Optoelectronics, South China University of Technology,  Guangzhou 510640, China}
\author{Xiaomin Wei}
\affiliation{School of Physics and Optoelectronics, South China University of Technology,  Guangzhou 510640, China}
\author{Zhongmin Yang}
\email[E-mail: ]{yangzm@scut.edu.cn}
\affiliation{School of Physics and Optoelectronics, South China University of Technology, Guangzhou 510640, China}
\affiliation{Research Institute of Future Technology, South China Normal University, Guangzhou 510006, China}
\affiliation{State Key Laboratory of Luminescent Materials and Devices and Institute of Optical Communication Materials, South China University of Technology, Guangzhou 510640, China}
\author{Tao Liu}
\email[E-mail: ]{liutao0716@scut.edu.cn}
\affiliation{School of Physics and Optoelectronics, South China University of Technology,  Guangzhou 510640, China}
\author{Franco Nori}
\affiliation{Center for Quantum Computing, RIKEN, Wakoshi, Saitama 351-0198, Japan}
\affiliation{Department of Physics, University of Michigan, Ann Arbor, Michigan 48109-1040, USA}

\date{{\small \today}}


\begin{abstract}
	The non-Hermitian skin effect (NHSE), characterized by the accumulation of a macroscopic number of bulk states at system boundaries, is a hallmark of non-Hermitian physics. However, in higher dimensions, achieving deterministic control over where skin modes accumulate remains a major challenge. Here, we propose a versatile route to program the skin-mode localization site in two-dimensional non-Hermitian lattices by combining disorder with a static electric field. While the electric field alone suppresses the NHSE in a clean system, the introduction of disorder induces transverse wave-packet transport perpendicular to the field. In nonreciprocal lattices, when the nonreciprocal hopping is misaligned with the electric field, the hopping component perpendicular to the field guides wave-packet propagation and produces boundary localization. By tuning the relative orientation between the electric field and the nonreciprocal hopping direction, the boundary localization position can be continuously and arbitrarily controlled. We further demonstrate distinct geometry-dependent manipulation of skin modes in reciprocal lattices, where controllable boundary localization emerges solely from the lattice geometry. Our results establish a robust and broadly applicable route to engineer boundary accumulation and directed transport along prescribed directions in two-dimensional non-Hermitian systems, enabling reconfigurable wave routing in classical platforms and programmable transport functionalities in quantum settings.
\end{abstract}
\maketitle

\section{Introduction}\label{Section1}

Non-Hermitian systems, described by Hamiltonians incorporating non-conservative processes such as gain and loss, have emerged as a fertile platform for uncovering unconventional physical phenomena \cite{RevModPhys.93.015005,Ashida2020}. By challenging conventional notions of symmetry, topology, and dynamics, they have spurred intense theoretical and experimental exploration in recent years \cite{Rotter2019,Naghiloo2019,Imhof2018,ShunyuYao2018,PhysRevLett.123.066404,PhysRevLett.125.126402, PhysRevLett.122.076801, PhysRevLett.121.026808,Leefmans2022,arXiv:1802.07964,  YaoarXiv:1804.04672, PhysRevLett.123.170401, PhysRevLett.123.206404,PhysRevLett.123.206404, PhysRevX.9.041015,9yj7-78sf, PhysRevLett.124.086801, PhysRevLett.127.196801,  PhysRevLett.129.093001, arXiv:2311.03777, vxgf-59xt,PhysRevLett.133.076502,Arkhipov2023,Lin2025,qrh6-vx64, Cai2025,Wang2025,q4nh-m1jh,vvrx-mljg,arxiv.2509.18828,PhysRevResearch.7.L012068,svww-ycws}. Among the most striking manifestations is the non-Hermitian skin effect (NHSE), where a macroscopic number of eigenstates accumulate at system boundaries \cite{ShunyuYao2018,PhysRevLett.123.066404,PhysRevLett.125.126402, PhysRevLett.122.076801, PhysRevLett.121.026808}, rooted in the point-gap topology of the complex eigenenergy spectrum. The NHSE gives rise to diverse physical phenomena, including the breakdown of bulk–boundary correspondence \cite{ShunyuYao2018,PhysRevLett.123.066404,PhysRevLett.125.126402}, directional amplification of excitations \cite{PhysRevB.103.L241408,Wanjura2020,PhysRevB.106.024301}, unconventional critical behavior \cite{Li2020,PhysRevB.108.L161409,PhysRevA.109.063329}, and the emergence of exotic photon–emitter dressed states \cite{PhysRevLett.129.223601}.

The NHSE, with its localization of skin modes, exhibits significant controllability through diverse physical parameters.  In one-dimensional systems, key control knobs include external electric fields \cite{PhysRevB.106.L161402}, impurity \cite{Li2021,PhysRevResearch.5.033058}, nonlinearity \cite{PhysRevLett.134.243805,vxgf-59xt,Cai2025}, and  disorder \cite{PhysRevLett.126.166801,arXiv:2311.03777,PhysRevLett.134.196302}. Extending to two dimensions, external magnetic fields provide an additional powerful handle, capable of either enhancing or suppressing the NHSE \cite{PhysRevLett.127.256402,PhysRevLett.131.116601,PhysRevB.106.L081402}. Despite this progress, existing approaches mainly tune whether the NHSE appears and how strongly modes localize. By contrast, the more demanding goal, deterministic, programmable control of the localization direction and the ultimate accumulation site in higher dimensions, remains largely open. This motivates a central question: Can skin-mode accumulation in a two-dimensional (2D) system be engineered to follow an arbitrary, designer-specified direction, and thereby be routed to a selected position along a given one-dimensional boundary? Achieving such spatially programmable control of the skin-mode destination would enable reconfigurable wave steering and transport across classical and quantum platforms, including directional signal routing, selective energy deposition, and robust propagation along prescribed directions in 2D media.

\begin{figure*}[!tb]
	\centering
	\includegraphics[width=18cm]{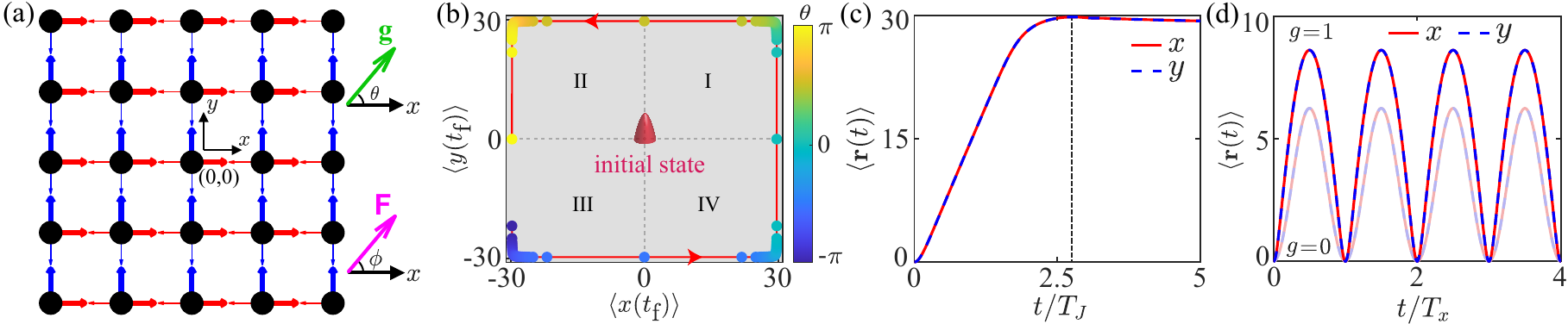}
	\caption{(a) Schematic of the 2D Hatano–Nelson model with nonreciprocal hopping, subject to a static electric field $\mathbf{F}$, oriented at angle $\phi$ (magenta arrow),  and random on-site disorder. The nonreciprocal hopping direction (green arrow) is encoded in the  vector $\mathbf{g} = (g\cos\theta,~g\sin\theta)$.  	  (b) Center-of-mass  $\langle x(t_\textrm{f}) \rangle$ and $\langle y(t_\textrm{f}) \rangle$, in the clean limit ($\xi=0$), at different $\theta \in (-\pi,\pi]$ (red arrows indicate increasing $\theta$) and   grouped into quadrants $\textrm{I}$--$\textrm{IV}$.   The initial Gaussian wave packet is   centered at $\mathbf{r}=(0,0)$ on a $61\times61$ lattice with $g/J=1$ and $F/J=0$, evolved up to $t_\textrm{f}=20T_J$ with $T_J=2\pi/J$.	(c) Time evolution of the center-of-mass $\langle \mathbf{r}(t) \rangle$ [red solid (blue dashed) line denotes $x$ ($y$) component]  in the clean limit ($\xi=0$), for $(g/J,\theta,F/J,\phi) = (1,\pi/4,0,0)$, with time in units of $T_J$. (d) $\langle \mathbf{r}(t) \rangle$ under a finite electric field  $F/J=0.8$ oriented at $\phi=\pi/4$, with time rescaled by  $T_x=2\pi/F_x$. }\label{Fig1}
\end{figure*}

In this paper, we realize deterministic control of 2D skin-mode boundary localization by combining random on-site disorder with a static electric field. In disordered Hermitian systems, such a field induces Anderson–Stark physics, including power-law localization  \cite{PhysRevB.33.780} in 1D and diffusion persisting over algebraically long time scales, ultimately leading to confinement within a finite region of the lattice \cite{PhysRevLett.101.190602}. By contrast, in 2D nonreciprocal non-Hermitian lattices, the interplay between disorder and the electric field unlocks a qualitatively different functionality. While the electric field alone typically suppresses the NHSE via Stark localization, we show that the disorder induces effective coupling among localized Wannier–Stark states, and when combined with nonreciprocal hopping, it rectifies the transverse dynamics into a net drift perpendicular to the field due to NHSE-induced Anderson delocalization along the transverse direction. This allows wave packets to be localized at arbitrary positions along the boundary by simply tuning the relative orientation of the field and the nonreciprocal hopping direction. Moreover, we show that analogous boundary-localization control is achievable even in reciprocal lattices, where NHSE typically depends sensitively on lattice geometry, thereby extending the scope of programmable steering beyond the conventional nonreciprocal setting.    Overall, our results reveal a synergistic interplay among the NHSE, disorder, and static electric fields that enables precise and programmable control over the boundary localization position, capabilities not available in previously studied non-Hermitian settings, and provides a simple and broadly applicable route to steer wave transport along designer-specified directions beyond the limits of Hermitian systems.

This article is organized as follows. In Sec.~\ref{Section2}, we introduce a two-dimensional extension of the Hatano–Nelson model under a static electric field and random on-site disorder, and show how their interplay enables flexible control of the boundary localization of skin modes. In Sec.~\ref{Section3}, we extend the discussion to reciprocal non-Hermitian lattices, where we explore geometry-dependent manipulation of skin modes under the combined effects of disorder and an external field. Finally, Sec.~\ref{Section4} provides a summary of the main results and an outlook for future work.

\section{Nonreciprocal model}\label{Section2}

We first consider a 2D extension of the Hatano–Nelson model \cite{PhysRevLett.77.570}, subject  to a  static electric field  and random on-site disorder, as shown in Fig.~\ref{Fig1}(a). This nonreciprocal model is defined on a square lattice, governed by the Hamiltonian
\begin{align}\label{Hamil}
	\hat{H} =   \sum_{\langle \mathbf{r},\mathbf{r^\prime}\rangle} J_{\mathbf{r},\mathbf{r^\prime}} \hat{c}_\mathbf{r}^\dagger \hat{c}_\mathbf{r^\prime} 
	+\sum_{\mathbf{r}} (\mathbf{F}\cdot\mathbf{r}) \hat{c}_\mathbf{r}^\dagger \hat{c}_\mathbf{r} +  \sum_{\mathbf{r}} \xi V_\mathbf{r}  \hat{c}_\mathbf{r}^\dagger \hat{c}_\mathbf{r},
\end{align}
where $\hat{c}_\mathbf{r}^\dagger$ and $\hat{c}_\mathbf{r}$  denote  the  creation and annihilation operators at the lattice site $\mathbf{r}=(x,y)$, and $\langle \mathbf{r},\mathbf{r^\prime}\rangle$ represents a pair of nearest-neighbor sites. The nonreciprocal hopping amplitudes along the $x$ and $y$ directions  are specified as 
\begin{align}\label{hopping}
	J_{(x\pm1,y),(x,y)} = J e^{\pm g_x},~~\textrm{and}~~J_{(x,y\pm1),(x,y)} = J e^{\pm g_y},
\end{align}
where  $g_x = g \cos \theta $ and $g_y = g \sin \theta $. The propagation direction  dictated by the NHSE  is thus controlled by the angle $\theta$ [see green arrow in Fig.~\ref{Fig1}(a)].   The external electric field term, $\mathbf{F}\cdot\mathbf{r}$, introduces a position-dependent potential along both the $x$ and $y$ directions, with the  field   $\mathbf{F}=(F_x,F_y)=F(\cos\phi, \sin\phi)$ oriented at the angle $\phi$ [see magenta arrow in Fig.~\ref{Fig1}(a)]. The term $\xi V_\mathbf{r}$ represents the random on-site potential with  $V_\mathbf{r} \in [-1/2,1/2]$ and disorder strength $\xi$.  

\subsection{Field-induced suppression of NHSE in pristine lattice}

Prior to addressing the combined impact of the electric field and disorder on skin-mode localization, we investigate the clean case ($\xi = 0$) to clarify how the electric field alone modifies the NHSE.

Starting from an initially-localized Gaussian wave packet $\ket{\Psi(0)} = \sum_{\mathbf{r}} \psi_{\mathbf{r}}(0) \ket{\mathbf{r}}$, with $\psi_{\mathbf{r}}(0) = e^{-(x^2+y^2)/\sigma^2}$ and   width $\sigma=2$ centered at $\mathbf{r}=(0,0)$, the normalized time-evolved state $\ket{\Psi(t)}$ is given by
\begin{align}\label{evolvedstate}
	\ket{\Psi(t)} = \frac{e^{-i\hat{H}t}\ket{\Psi(0)}}{\abs{\abs{e^{-i\hat{H}t}\ket{\Psi(0)}}}}.
\end{align}
We characterize its motion by tracing the center of mass 
\begin{align}\label{COM}
	\langle \mathbf{r}(t) \rangle = \bra{\Psi(t)} \hat{\mathbf{r}} \ket{\Psi(t)}.
\end{align}

In the absence of an electric field, the dynamics is dominated by the NHSE. Consequently, the wave packet localizes either along an edge (only for $\theta=0, \pm\pi/2, \pi$) or at a corner, as reflected in the long-time center-of-mass $\langle \mathbf{r}(t_\textrm{f}) \rangle$ shown in Fig.~\ref{Fig1}(b). A representative trajectory for $\theta = \pi/4$ is shown in Fig.~\ref{Fig1}(c), where the wave packet propagates toward the boundary and eventually becomes confined in the corner, without back-propagation, due to the NHSE.

A static electric field qualitatively alters the localization properties of the skin modes due to the emergence of field-induced Stark localization \cite{RevModPhys.34.645}. To investigate this effect, we analytically obtain the time-evolved state $\ket{\Psi(t)}$ by solving the Schr\"odinger equation $i \partial_t \ket{\Psi(t)}= \hat{H} \ket{\Psi(t)}$ for the disorder-free Hamiltonian $\hat{H}$ in Eq.~(\ref{Hamil}).  As derived in Appendix \ref{AppendixA}, the time-evolved state is expressed as 
\begin{align}\label{FinalResult2}
	\ket{\Psi(t)} = \frac{1}{\sqrt{\mathcal{N}}} \sum_{\mathbf{r}} \psi_\mathbf{r}(t) \ket{\mathbf{r}}, 
\end{align}
with
\begin{align}\label{FinalResult}
	\psi_{\mathbf{r}}(t) = ~& e^{\mathbf{g}\cdot\mathbf{r}} \sum_{\mathbf{r}^\prime} (-i)^{(x^\prime-x)+(y^\prime-y)} \mathcal{J}_{x-x^\prime}(z_x) \mathcal{J}_{y-y^\prime}(z_y) \nonumber \\
	& \times e^{-i [(x^\prime+x) \varphi_x + (y^\prime+y) \varphi_y]}  \psi_{\mathbf{r}^\prime}(0),
\end{align}
where $\varphi_\alpha = F_\alpha t /2$ ($\alpha=x,y$),   $z_\alpha = 4 J \abs{\sin (F_\alpha t/2)/ F_\alpha}$,   $\mathcal{J}_{n}(z)$ denotes the Bessel function of the first kind, and $\mathcal{N}$ is the normalization coefficient. 

According to Eq.~(\ref{FinalResult}), when $z_\alpha=0$, the Bessel function reduce to
$\mathcal J_{x-x'}(0)\mathcal J_{y-y'}(0)=\delta_{x,x'}\delta_{y,y'}$, so that only $\mathbf r=\mathbf r'$ contributes to the sum. Hence, at these special times, the wave packet revives and returns
to its initial position. More broadly, a static electric field produces Wannier--Stark localization \cite{RevModPhys.34.645}, which confines
the dynamics in real space and suppresses long-range intersite transport even for strong nonreciprocal
hopping. As a result, the electric
field suppresses the NHSE by preventing the wave packet from developing a net drift and accumulating
at the boundary. Instead, the system displays  modified Bloch oscillations  shaped by the
nonreciprocal hopping: rather than being advected unidirectionally toward an edge, a bulk-initialized
wave packet stays Stark-localized and undergoes bounded oscillatory motion within the bulk, as shown
in Fig.~\ref{Fig1}(d).  

\begin{figure*}[!tb]
	\centering
	\includegraphics[width=18cm]{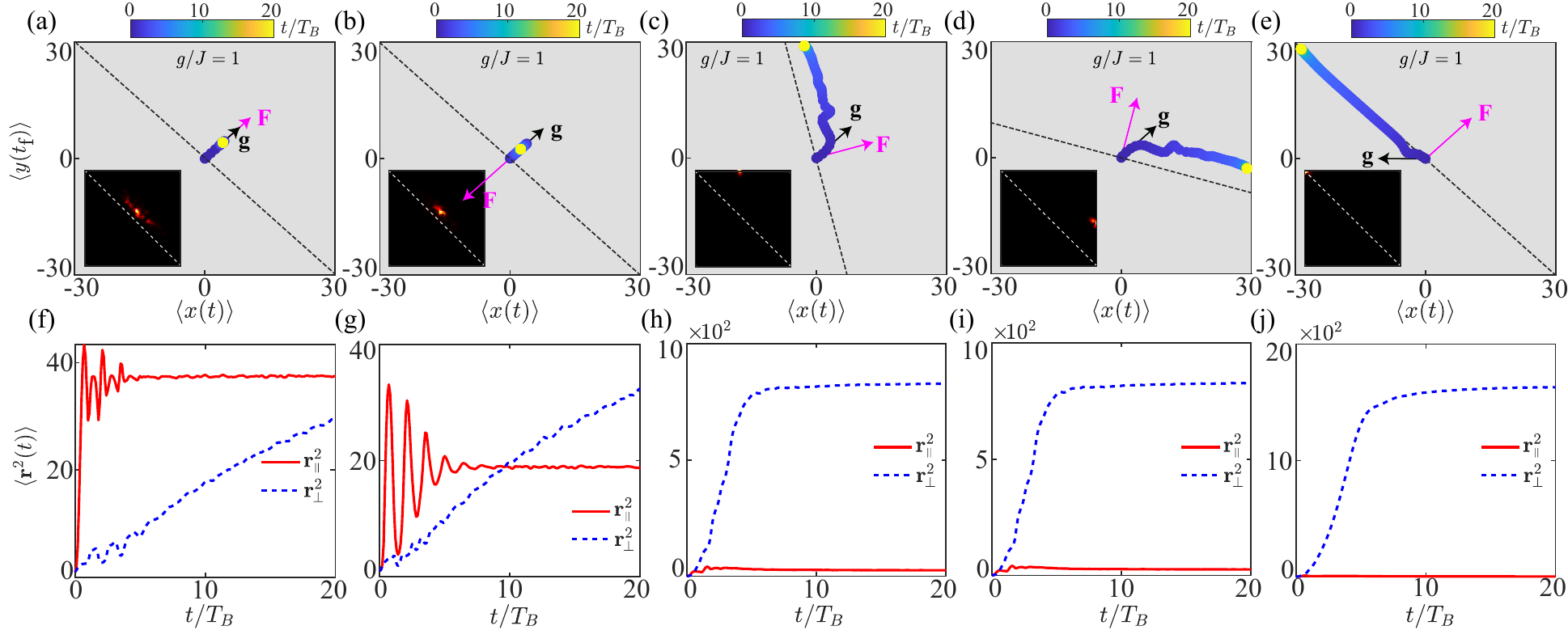}
	\caption{(a–e) Time-evolution trajectory of   the center of mass, $\langle x(t) \rangle$ and $\langle y(t) \rangle$, for an initially Gaussian wave packet centered at the origin, under an electric field with $F/J = 1.5$ and random on-site  disorder $\xi/J = 1.0$.   The nonreciprocal hopping strength is fixed at $g/J = 1$.   The nonreciprocity direction and   the electric-field orientation are chosen as for $(\theta,\phi) = (\pi/4,\pi/4)$ (a), $(\pi/4,-3\pi/4)$ (b), $(\pi/4,\pi/12)$ (c), $(\pi/4,5\pi/12)$ (d), and $(\pi,\pi/4)$ (e). Circle color encodes time in units of the Bloch period $T_B = 2\pi/F$. Insets show the corresponding probability density distributions after long-time evolution. (f–j) Corresponding time-resolved second moment $\langle \mathbf{r}^2(t) \rangle$, where red solid and blue dashed lines denote components parallel and perpendicular to the electric field, respectively. All results are averaged over 1000 disorder realizations.}\label{Fig2}
\end{figure*}

\subsection{Interplay of electric field and disorder}

We have demonstrated that the static electric field  suppresses the NHSE in the 2D nonreciprocal lattice. We now extend our study to include random on-site disorder. In non-Hermitian systems without an electric field, disorder competes with the NHSE, leading to a skin-Anderson transition and distinct dynamical phenomena \cite{PhysRevLett.77.570,PhysRevX.8.031079,PhysRevLett.134.176301,arXiv:2507.14523,wsmq-kmq9}. Here, we show that, in the  2D nonreciprocal systems, the interplay between disorder and a static electric field enables control over the propagation direction of skin modes, allowing precise manipulation of their localization position along the boundary. 

To intuitively understand the role of disorder in the directional hopping under both nonreciprocal hopping and a static electric field, we rewrite the Hamiltonian in Eq.~\eqref{Hamil} in the biorthogonal modified Wannier–Stark basis $\ket{\bar{\Psi}_{m,n}}$ and $\ket{\Psi_{m,n}}$ in the thermodynamic limit (see details in Appendix \ref{AppendixB}) as
\begin{align}\label{fullH}
	\hat{H} = & \sum_{m,n} \left(F_x m + F_y n\right) \ket{\Psi_{m,n}} \bra{\bar{\Psi}_{m,n}} \nonumber \\
	& + \sum_{m,n}\sum_{m^\prime,n^\prime} V_{(m,n),(m^\prime,n^\prime)}
	\ket{\Psi_{m,n}} \bra{\bar{\Psi}_{m^\prime,n^\prime}},
\end{align}
where the disorder-induced couplings are given by
\begin{align}\label{Velements2}
	V_{(m,n),(m^\prime,n^\prime)}  =~ & \xi \sum_{x,y} V_{x,y}
	\mathcal{J}_{x-m} (\gamma_x)
	\mathcal{J}_{y-n} (\gamma_y) \nonumber \\
	& \times \mathcal{J}_{x-m^\prime}(\gamma_x)
	\mathcal{J}_{y-n^\prime}(\gamma_y),
\end{align}
with $\gamma_\alpha = -2J/F_\alpha$ ($\alpha = x,y$). The right and left Wannier–Stark states are
\begin{align}\label{WannierStateNHR11}
	\ket{\Psi_{m,n}} = \sum_{x,y} \exp(g_x x + g_y y) \mathcal{J}_{x-m}(\gamma_x) \mathcal{J}_{y-n}(\gamma_y) \ket{x,y},
\end{align}
\begin{align}\label{WannierStateNHL22}
	\ket{\bar{\Psi}_{m,n}} = \sum_{x,y} \exp(-g_x x - g_y y) \mathcal{J}_{x-m}(\gamma_x) \mathcal{J}_{y-n}(\gamma_y) \ket{x,y}.
\end{align}

As indicated by the effective coupling in Eq.~(\ref{Velements2}), random on-site disorder facilitates hopping between localized Wannier–Stark states, thereby effectively opening new transport channels. Meanwhile, the nonreciprocal hopping introduces directional bias in this transport, as
indicated by our numerical results below.

We now numerically examine the dynamical evolution of center of mass and the spreading dynamics of a Gaussian wave packet centered at the origin under an electric field with $F/J = 1.5$ and random on-site  disorder $\xi/J = 1.0$.  The wave spreading is quantified by the  second moment of wave-packet distribution  
\begin{align}
\langle \mathbf{r}^2(t) \rangle = \bra{\Psi(t)} \hat{\mathbf{r}}^2 \ket{\Psi(t)}.  
\end{align}

We define the unit vectors parallel and perpendicular to the electric field as $\mathbf{r}_\parallel = (\cos\phi, \sin\phi)$, and  $\mathbf{r}_\perp = (-\sin\phi, \cos\phi)$, and evaluate the corresponding parallel and perpendicular components of the second moment $\langle \mathbf{r}^2(t) \rangle$. The second moment scales as $\langle \mathbf{r}^2(t) \rangle \sim t^{\beta}$, with $\beta$ determining the transport regime: subdiffusive ($0 < \beta < 1$), diffusive ($\beta = 1$), superdiffusive ($1 < \beta < 2$), and ballistic ($\beta = 2$).

We consider that the nonreciprocal system is subject to the static electric field and random on-site disorder. Specifically, we consider two representative cases, $\theta = \phi$ and $\theta = \pi - \phi$, where the direction defined by the vector $\mathbf{g}$, which sets the propagation direction associated with the NHSE, is either parallel [Fig.~\ref{Fig2}(a)] or antiparallel [Fig.~\ref{Fig2}(b)] to the electric field $\mathbf{F}$. We calculate the time evolution of the wave-packet center of mass, $\langle x(t) \rangle$ and $\langle y(t) \rangle$, for an initially Gaussian wave packet centered at the origin. As shown in Figs.~\ref{Fig2}(a,b), the wave packet remains localized along the field direction, with its center exhibiting a slight displacement from the origin induced by the nonreciprocal hopping. In contrast, along the direction perpendicular to the field, the wave packet spreads diffusively over a finite time window, as indicated by the time-dependent second moment $\langle \mathbf{r}^2(t) \rangle$ in Figs.~\ref{Fig2}(f, g). In this configuration, the wave packet shows no net directed drift toward the boundary, but  is confined within a finite region of the lattice.

When the direction of $\mathbf{g}$ deviates from being parallel or antiparallel to $\mathbf{F}$, qualitatively new dynamics emerge. As shown in Figs.~\ref{Fig2}(c–e), the trajectories of $\langle x(t) \rangle$ and $\langle y(t) \rangle$ reveal that the component of $\mathbf{g}$ parallel to the electric field induces only a slight shift of the wave-packet center from the origin and does not contribute to long-time transport. In contrast, the component perpendicular to the field drives asymmetric diffusive spreading toward the boundary due to NHSE-induced Anderson delocalization along the transverse direction. This process causes the wave packet to drift gradually and eventually become localized at the system boundary [see Figs.~\ref{Fig2}(c–e) and the corresponding second moment $\langle \mathbf{r}^2(t) \rangle$ in Figs.~\ref{Fig2}(h–j)]. The transport direction of the wave packet is primarily determined by the component of $\mathbf{g}$ perpendicular to the electric field, indicating that the final localization position can be tuned by the field orientation. This behavior highlights the remarkable interplay among the NHSE, disorder, and electric field in controlling the transport of wave packets in 2D nonreciprocal non-Hermitian systems. 

\begin{figure*}[!tb]
	\centering
	\includegraphics[width=18cm]{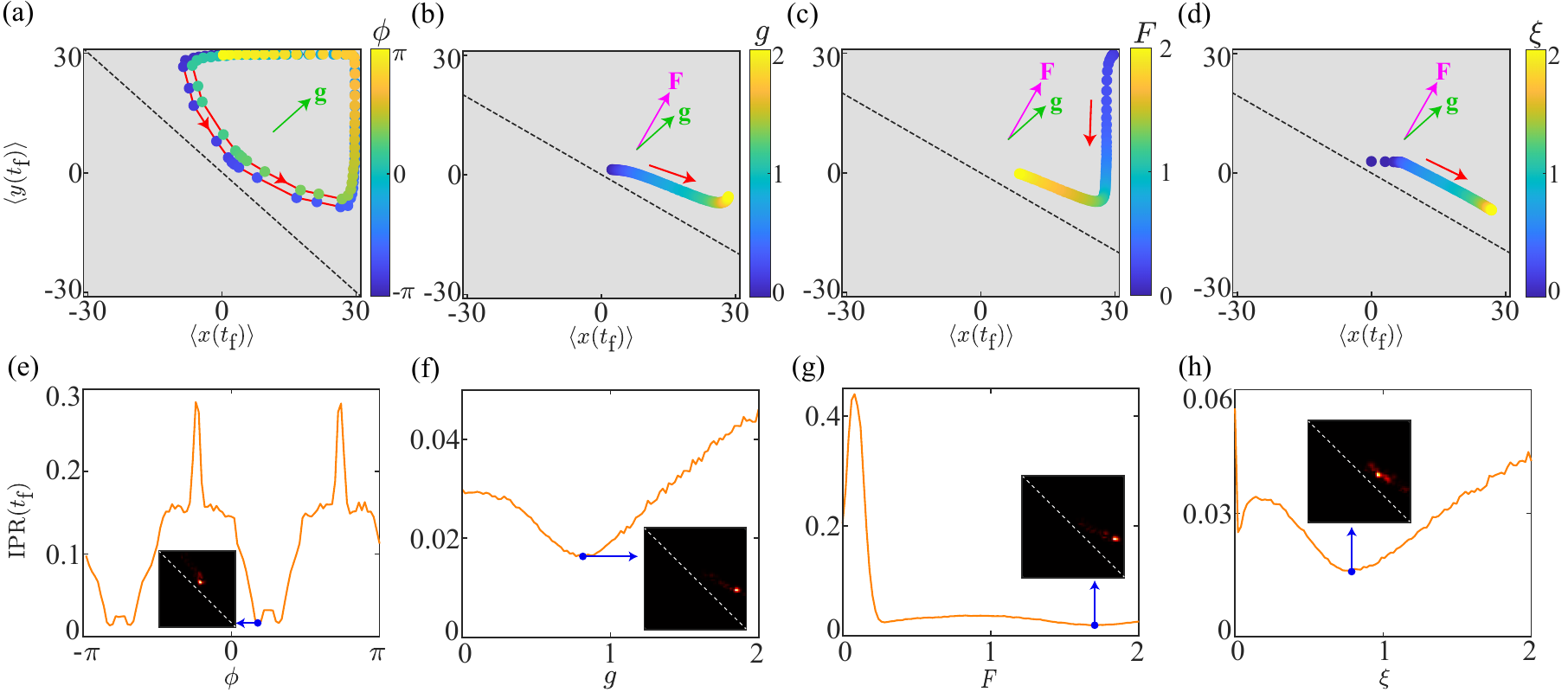}
	\caption{(a-d) Trajectories of the center of mass, $\langle x(t_\mathrm{f})\rangle$ and $\langle y(t_\mathrm{f})\rangle$, of an origin-centered Gaussian wave packet after long-time evolution under nonreciprocal hopping, electric field, and random on-site disorder, versus  field orientation  $\phi$ (a), nonreciprocity strength $g$ (b), field strength  $F$ (c), and disorder strength  $\xi$ (d).    Insets mark the NHSE (green) and field (magenta) orientations. Colors denote the varied parameter (increasing along red arrows). Parameters used are $g/J=1.0$, $F/J=1.5$, $\xi/J=1.0$ and $t_\textrm{f} = 20 T_B$, with $\theta=\pi/4$ in (a), and $(\theta, \phi)=(\pi/4,\pi/3)$ in (b-d).  (e-h) IPR of the evolved states at time $t_\textrm{f}$. The insets show the corresponding probability density distributions for the  states with the smallest IPR.  All results  are averaged over 1000 disorder realizations.   }\label{Fig3}
\end{figure*}

\subsection{Versatile control of mode localization}

We now systematically investigate how the wave-packet transport and localization direction can be tuned via the combined effects of the nonreciprocal hopping, disorder, and electric field. 

Figure~\ref{Fig3}(a–d) shows the long-time center-of-mass  $\langle \mathbf{r}(t_\mathrm{f}) \rangle$ of an initially localized Gaussian wave packet at the origin, plotted as a function of the field orientation $\phi$ (a), nonreciprocity strength $g$ (b), field strength $F$ (c), and disorder strength $\xi$ (d). When the nonreciprocal hopping parameter $\mathbf{g}$, with $\theta = \pi/4$, is fixed, varying the direction of the applied electric field allows the wave packet to localize at arbitrary positions along the boundary of the upper-right quadrant [see Fig.~\ref{Fig3}(a)].    

Furthermore, when the directions of $\mathbf{F}$ and $\mathbf{g}$ are fixed, numerical simulations reveal that boundary localization occurs only when the nonreciprocity strength $g$ [see Fig.~\ref{Fig3}(b)], field strength $F$ [see Fig.~\ref{Fig3}(c)], or disorder strength $\xi$ [see Fig.~\ref{Fig3}(d)] are tuned to appropriate values. 
In addition, the ultra-long-time evolution results further confirm that these wave packets remain well localized at the boundary (See details in Appendix \ref{AppendixC}).

To further verify this boundary localization, we analyze the inverse participation ratio (IPR), defined as
\begin{align}\label{fullH2SM001}
	\mathrm{IPR}(t) = \sum_{\mathbf{r}} |\langle \mathbf{r} | \Psi(t) \rangle|^4. 
\end{align}
For an extended eigenstate $\Psi(t_f)$, $\mathrm{IPR}(t_f) \simeq 1/(L_x L_y)$, which vanishes as $N \to \infty$, whereas for a state localized on $M \ll L_x L_y$ sites, $\mathrm{IPR}(t_\textrm{f}) \simeq 1/M$ and thus remains finite in the thermodynamic limit.

In Figs.~\ref{Fig3}(e–h), we plot the long-time IPR as a function of $\phi$ (e), $g$ (f), $F$ (g), and $\xi$ (h). The insets show the corresponding probability-density distributions for the states with the smallest IPR. These results confirm that the wave packet becomes localized after long-time evolution, consistent with the conclusions drawn from the long-time center-of-mass behavior $\langle \mathbf{r}(t_\mathrm{f}) \rangle$.

These results highlight the synergistic interplay among the nonreciprocal hopping, disorder, and static electric field, which together enable precise control over the localization position of the wave packet in 2D non-Hermitian systems, a capability unattainable in conventional settings. This interplay provides a versatile mechanism for engineering tunable boundary accumulation in realistic and disordered systems. Note that, while our analysis focuses on non-Hermitian Hamiltonians, we  demonstrate that similar dynamics can arise in open quantum systems (see Appendix \ref{AppendixD}), where coherent–dissipative interplay generates effective nonreciprocal hopping.

\begin{figure*}[!tb]
	\centering
	\includegraphics[width=16cm]{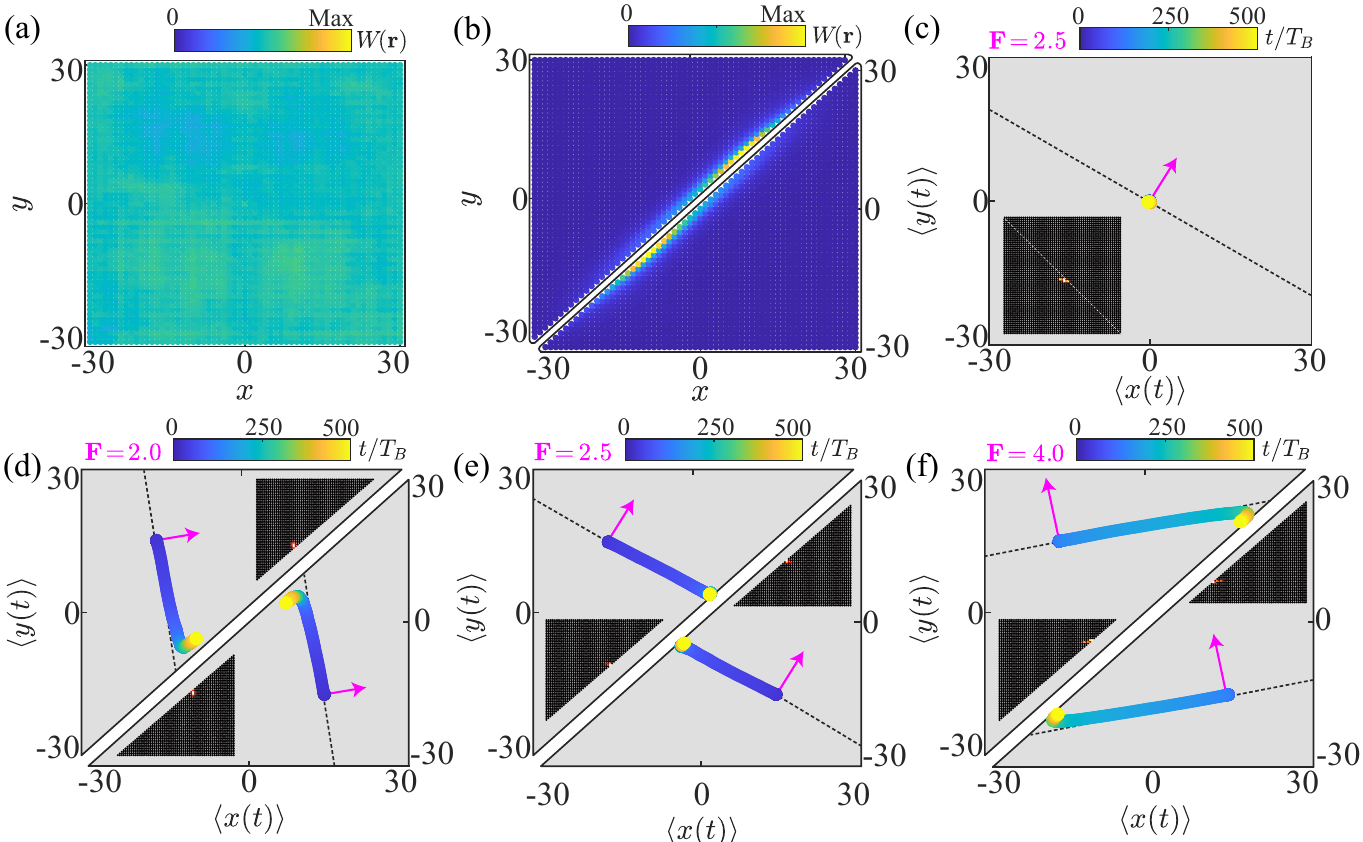}
	\caption{(a,b) Spatial distributions of eigenstates $W(\mathbf{r})$ for  the square geometry (a) and  the up- and down-triangle geometries (b), with color bars indicating intensity. (c–f) Time-evolution trajectories of the center of mass, $\langle x(t)\rangle$ and $\langle y(t)\rangle$, for an initial wave packet located at the site marked by the magenta arrow. Results are shown for (c) the square geometry with $(F/J,\phi) = (2.5,\pi/3)$ and for (d–f) triangle geometries with (d) $(2.0,\pi/24)$, (e) $(2.5,\pi/3)$, and (f) $(4.0,7\pi/12)$. Circle color denotes time in units of $T_B = 2\pi/J$. Insets display the probability density distributions after a long-time evolution. Parameters: $J=1$, $J_x = 1.0J$, $J_y = 0.5iJ$, $\xi/J = 0.5$, and $L_x \times L_y = 61 \times 61$.  All results are averaged over 1000 disorder realizations.}\label{FigS3}
\end{figure*}

\section{Geometry-dependent manipulation of skin modes in reciprocal lattice}\label{Section3}

We have revealed the versatile control of skin-mode localization in nonreciprocal systems via the electric field and disorder. Beyond the nonreciprocal case, the NHSE can also arise in reciprocal systems \cite{Zhang2022SM, PhysRevLett.131.207201, Zhou2023SM}. Unlike in nonreciprocal 2D systems, the NHSE in reciprocal lattices is inherently geometry-dependent \cite{Zhang2022SM}: the interplay between lattice symmetry and spectral reciprocity causes skin modes to accumulate at specific crystalline interfaces under open boundaries. We now demonstrate the geometry-dependent manipulation of skin modes in a reciprocal lattice via the electric field and disorder.

We consider a 2D reciprocal non-Hermitian lattice subject to disorder and an external electric field, described by   Hamiltonian
\begin{align}\label{HamilgSM}
	\hat{H}_\textrm{R} =   &\sum_{x,y} \left[ J_x (\hat{c}^\dagger_{x+1,y} \hat{c}_{x,y} + \text{H.c.}) +   J_y (\hat{c}^\dagger_{x,y+1} \hat{c}_{x,y} + \text{H.c.})\right] \nonumber \\
	&+\sum_{x,y} \left[(F_x x + F_y y) \hat{c}^\dagger_{x,y} \hat{c}_{x,y} + \xi   V_{x,y} \hat{c}^\dagger_{x,y} \hat{c}_{x,y} \right],
\end{align}
where $J_x$ and $J_y$ are complex coupling strengths along the $x$ and $y$ directions. 

Without disorder and an electric field, the system exhibits a geometry-dependent skin effect \cite{Zhang2022SM}, as shown in Figs.~\ref{FigS3}(a,b) for square and triangular geometries. The spatial distribution of all eigenstates is characterized by $W(\mathbf{r}) = \frac{1}{L} \sum_{n=1}^{L} \abs{\psi_n(\mathbf{r})}^2$, where $\psi_n(\mathbf{r})$ are normalized right eigenstates and $L$ is the total number of eigenstates. While the square geometry hosts extended eigenstates, up- and down-triangle geometries exhibit localization along the slanted edge.

Upon introducing both disorder and an electric field, we compute the time-evolution of the center of mass, $\langle x(t)\rangle$ and $\langle y(t)\rangle$, for an initially localized wave packet [Figs.~\ref{FigS3}(c–f)]. In the square geometry, the wave packet remains confined along the electric-field direction and spreads diffusively within a limited time in the transverse direction [Fig.~\ref{FigS3}(c)]. By contrast, in up- and down-triangle geometries under various field orientations, the wave packet drifts and ultimately accumulates along the slanted edge perpendicular to the field [Figs.~\ref{FigS3}(d–f)]. These results demonstrate that the interplay among lattice geometry, disorder, and electric field enables geometry-dependent manipulation of wave-packet dynamics and the localization position of skin modes in 2D reciprocal non-Hermitian systems.

\section{Conclusion}\label{Section4}

To summarize, we have demonstrated arbitrary manipulation of the NHSE and versatile control over the localization position of skin modes by utilizing random on-site disorder and a static electric field in 2D lattices. We show that the electric field suppresses the NHSE in clean lattices. Remarkably, in nonreciprocal lattices, the interplay between nonreciprocal hopping, disorder, and the electric field allows precise control over the direction of wave-packet localization, enabling boundary accumulation at arbitrary positions by tuning the electric-field orientation and nonreciprocal hopping. We further uncover a distinct, geometry-dependent route to control skin modes in reciprocal lattices: whereas boundary localization is usually dictated by the lattice geometry, here it can be reconfigured through the combined action of disorder and a static electric field, enabling geometry-dependent and programmable control of skin-mode localization position.   

Overall, our results establish a unified framework for understanding and manipulating the combined effects of NHSE, external fields, and disorder, with direct implications for wave control and transport along prescribed directions across classical and quantum platforms, for example, directional signal amplification, reconfigurable wave steering, and robust energy delivery beyond the constraints of Hermitian systems. The proposed mechanism is readily implementable in classical platforms \cite{Weidemann2020,PhysRevLett.128.223903,Zou2021,Zhou2023,Zheng2024} and  ultracold-atom systems \cite{PhysRevLett.129.070401,Zhao2025}, where NHSEs have already been observed. In photonic lattices, the effective electric field can be engineered via light intensity-controlled refractive-index gradients \cite{PhysRevLett.96.053903} or femtosecond laser direct-written waveguide arrays \cite{Corrielli2013}, while in ultracold atoms it can be realized using tilted optical lattices \cite{PhysRevLett.111.185301,Scherg2021}.

Beyond near-term experimental tests, our work opens multiple directions for exploring non-Hermitian physics enabled by programmable skin-mode control, including the roles of different disorder classes, nonlinearity, many-body interactions, and quasicrystalline geometries, as well as device-oriented applications such as directional wave amplification and adaptive wave routing along designer-specified directions.

\begin{acknowledgments}
	T.L. acknowledges the support from the Guangdong Provincial Quantum Science Strategic Initiative (Grant No.~GDZX2505004),  National Natural Science Foundation of China (Grant No.~12274142), the Key Program of the National Natural Science Foundation of China (Grant No.~62434009),  Introduced Innovative Team Project of Guangdong Pearl River Talents Program (Grant No.~2021ZT09Z109). Y.R.Z. is supported in part by: the National Natural Science Foundation of China (Grant No.~12475017), the	Natural Science Foundation of Guangdong Province (Grant	No.~2024A1515010398), and the Startup Grant of South	China University of Technology (Grant No.~20240061). F.N. is supported in part by: the Japan Science and Technology Agency (JST) [via the CREST Quantum Frontiers program Grant No. JPMJCR24I2, the Quantum Leap Flagship Program (Q-LEAP), and the Moonshot R$\&$D Grant No.~JPMJMS2061].
\end{acknowledgments}

\section*{Data availability}
The data that support the findings of this article are not publicly available. The data are available from the authors upon reasonable request.

\appendix
\section{Analytical dynamics in the absence of disorder}\label{AppendixA}
In the absence of disorder, the evolving dynamics of the two-dimensional Hatano–Nelson model under a static electric field can be analytically solved. The disorder-free Hamiltonian is given by 
\begin{align}\label{HamilSM}
	\hat{H}_\textrm{f} =   \sum_{\langle \mathbf{r},\mathbf{r^\prime}\rangle} J_{\mathbf{r},\mathbf{r}^\prime} \hat{c}^\dagger_{\mathbf{r}} \hat{c}_{\mathbf{r}^\prime}
	+\sum_{\mathbf{r}} (\mathbf{F}\cdot\mathbf{r}) \hat{c}^\dagger_{\mathbf{r}} \hat{c}_{\mathbf{r}}.
\end{align}

The hopping asymmetry  can be gauged away via the  similarity transformation
\begin{align}\label{SSM}
	\hat{S} = \exp \left[\sum_{x,y} -(g_x x + g_y y ) \hat{c}^\dagger_{x,y} \hat{c}_{x,y}\right],
\end{align}
which acts as  
\begin{align}\label{ScSM}
	\hat{S} \hat{c}_{x,y} \hat{S}^{-1} = \exp(g_x x + g_y y ) \hat{c}_{x,y},.
\end{align}
\begin{align}\label{ScSM2}
	\hat{S} \hat{c}^\dagger_{x,y} \hat{S}^{-1} = \exp[-(g_x x + g_y y)] \hat{c}^\dagger_{x,y}.
\end{align}

Under this transformation, the transformed Hamiltonian $\hat{H}^\prime  = \hat{S} \hat{H}_\textrm{f} \hat{S}^{-1}$ becomes Hermitian, 
\begin{align}\label{HamilTSM}
	\hat{H}^\prime = J \sum_{\mathbf{r}} \sum_{\boldsymbol{\delta} \in \{\hat{\mathbf{x}},\hat{\mathbf{y}}\}} (\ket{\mathbf{r}+\boldsymbol{\delta}} \bra{\mathbf{r}} + \textrm{H.c.}) +\sum_{\mathbf{r}} (\mathbf{F}\cdot\mathbf{r}) \ket{\mathbf{r}} \bra{\mathbf{r}}.
\end{align}

The time-dependent Schr\"odinger equation 
\begin{align}\label{SchM}
	i \partial_t \ket{\Psi(t)}= \hat{H}_\textrm{f} \ket{\Psi(t)},
\end{align}
is therefore mapped to
\begin{align}\label{TDSESM}
	i\partial_t\ket{\Phi(t)}=\hat{H}^\prime\ket{\Phi(t)},
\end{align}
with the relation
\begin{align}\label{stateRSM}
	\ket{\Psi(t)}=\hat{S}^{-1}\ket{\Phi(t)}.
\end{align}

To obtain the exact wave-packet dynamics $\ket{\Phi(t)}$ in Eq.~(\ref{TDSESM}), we expand  the state in the real-space basis,
\begin{align}\label{StateSM}
	\ket{\Phi(t)} = \sum_{\mathbf{r}} \phi_{\mathbf{r}}(t) \ket{\mathbf{r}}.
\end{align}
The amplitudes $\phi_{\mathbf{r}}(t)$ satisfy
\begin{align}\label{SequationSM}
	i \frac{d}{dt} \phi_{\mathbf{r}}(t) = J \sum_{\boldsymbol{\delta} = \hat{\mathbf{x}},\hat{\mathbf{y}}} [\phi_{\mathbf{r}+\boldsymbol{\delta}}(t) + \phi_{\mathbf{r}-\boldsymbol{\delta}}(t)] + (\mathbf{F}\cdot\mathbf{r}) \phi_{\mathbf{r}}(t).
\end{align}

Next, we introduce the rotating-frame transformation
\begin{align}\label{rotatedSM}
	\phi_{\mathbf{r}}(t) = e^{-i\mathbf{F}\cdot\mathbf{r} t} C_{\mathbf{r}}(t),
\end{align}
which removes the linear potential term and yields
\begin{align}\label{Sequation2SM}
	i \frac{d}{dt} C_{x,y}(t) = &  J \sum_{x,y} \left[e^{-i F_x t}  C_{x+1,y}(t) + e^{-i F_y t}  C_{x,y+1}(t) \right. \notag \\
	& \left. ~~~~ + e^{i F_x t}  C_{x-1,y}(t) + e^{i F_y t}  C_{x,y-1}(t)\right].
\end{align}

We then perform the discrete Fourier transform
\begin{align}\label{FourierSM}
	C_{\mathbf{k}}(t) = \frac{1}{\sqrt{N}} \sum_{\mathbf{r}} e^{-i \mathbf{k} \cdot \mathbf{r}} C_{\mathbf{r}}(t),
\end{align}
where $N$ is the total number of lattice sites. The evolution equation in momentum space becomes
\begin{align}\label{equation2SM}
	\frac{d}{dt} C_{\mathbf{k}}(t) = -2 iJ \left[\cos (k_x-F_x t) + \cos (k_y-F_y t)\right]  C_{\mathbf{k}}(t).
\end{align}
Its solution can be obtained by integrating $t$ over two sides given by
\begin{align}\label{DiffResult2SM}
	C_{\mathbf{k}}(t) & = C_{\mathbf{k}}(0) \exp(-2iJ \int_0^t \left[\sum_{\alpha\in\{x,y\}} \cos(k_\alpha - F_\alpha t^\prime) \right] d t^\prime).
\end{align}

Using the integrals  
\begin{align}\label{FuncSM}
	\mathcal{U}_\alpha(t) = \int_0^t \cos (F_\alpha t^\prime) dt^\prime = \frac{\sin (F_\alpha t)}{F_\alpha},
\end{align}
and
\begin{align}\label{Func2SM}
	\mathcal{V}_\alpha (t) = \int_0^t \sin (F_\alpha t^\prime) dt^\prime = \frac{1-\cos (F_\alpha t)}{F_\alpha},
\end{align}
Equation~(\ref{DiffResult2SM}) can then be rewritten as
\begin{align}\label{DiffResult3SM}
	C_{\mathbf{k}}(t) & = C_{\mathbf{k}}(0) \exp\left[-2iJ \sum_{\alpha=x,y} (\sin k_\alpha \mathcal{V}_\alpha(t) + \cos k_\alpha \mathcal{U}_\alpha(t))\right].
\end{align}

Defining the amplitude and phase
\begin{align}\label{DefSM}
	A_\alpha(t) = \sqrt{\mathcal{U}_\alpha^2 + \mathcal{V}_\alpha^2} = \frac{2 \abs{\sin (F_\alpha t/2)}}{\abs{F_\alpha}},
\end{align}
\begin{align}\label{Def2SM}
	\varphi_\alpha(t) = \arctan  (\frac{\mathcal{V}_\alpha}{\mathcal{U}_\alpha}) = \frac{F_\alpha t}{2},
\end{align}
the solution takes the compact form
\begin{align}\label{DiffResult4SM}
	C_{\mathbf{k}}(t) & = C_{\mathbf{k}}(0) \prod_{\alpha=x,y} \exp[-2iJ A_\alpha \cos (k_\alpha - \varphi_\alpha)].
\end{align}

Using the Jacobi–Anger expansion
\begin{align}\label{BesselSM}
	\exp(-iz \cos \theta ) = \sum_{n=-\infty}^{\infty} (-i)^n \mathcal{J}_n(z) e^{in\theta},
\end{align}
with the Bessel function  $\mathcal{J}_n(z)$ of the first kind, and performing a Fourier inversion, we derive the exact real-space amplitude
\begin{align}\label{FinalResultSM}
	C_{\mathbf{r}}(t) = & \sum_{\mathbf{r}^\prime} (-i)^{(x^\prime-x)+(y^\prime-y)} \mathcal{J}_{x^\prime-x}(z_x) \mathcal{J}_{y^\prime-y}(z_y) \notag \\
	& ~~\times \exp(-i [(x^\prime-x) \varphi_x+ (y^\prime-y) \varphi_y]) C_{\mathbf{r}^\prime}(0),
\end{align}
where $z_\alpha(t) = 2JA_\alpha(t) = 4J\abs{\sin (F_\alpha t/2)/F_\alpha}$.

Restoring the original frame, we have
\begin{align}\label{FinalResult1SM}
	\phi_{\mathbf{r}}(t) = & \sum_{\mathbf{r}^\prime} (-i)^{(x^\prime-x)+(y^\prime-y)} \mathcal{J}_{x-x^\prime}(z_x) \mathcal{J}_{y-y^\prime}(z_y) \notag \\
	& ~~\times \exp(-i [(x^\prime+x) \varphi_x + (y^\prime+y) \varphi_y])  \phi_{\mathbf{r}^\prime}(0),
\end{align}
where $\phi_{\mathbf{r}^\prime}(0) = C_{\mathbf{r}^\prime}(0)$ for $t=0$. 

Finally, applying the inverse similarity transformation, the evolving state of the original non-Hermitian Hamiltonian is  
\begin{align}\label{FinalResult2SM}
	\ket{\Psi(t)}= \frac{1}{\sqrt{\mathcal{N}}} \sum_{\mathbf{r}}\psi_{\mathbf{r}}(t)\ket{\mathbf{r}},
\end{align}
with
\begin{align}\label{FinalResult1SM2}
	\psi_{\mathbf{r}}(t)= & e^{\mathbf{g}\cdot\mathbf{r}} \sum_{\mathbf{r}^\prime} (-i)^{(x^\prime-x)+(y^\prime-y)} \mathcal{J}_{x-x^\prime}(z_x) \mathcal{J}_{y-y^\prime}(z_y) \notag \\
	& ~~\times \exp(-i [(x^\prime+x) \varphi_x + (y^\prime+y) \varphi_y])  \phi_{\mathbf{r}^\prime}(0),
\end{align}
where $\varphi_\alpha  = F_\alpha t/2$ ($\alpha=x,y$),   $z_\alpha  = 4J\abs{\sin (F_\alpha t/2)/F_\alpha}$, and $\mathcal{N}$ being normalized coefficient.  

According to Eq.~(\ref{FinalResult1SM2}), the external electric field suppresses the non-Hermitian skin effect (NHSE), counteracting the wave packet's localization at the boundary despite strong nonreciprocal hopping. This suppression transforms the dynamics into modified Bloch oscillations due to nonreciprocal hopping. Instead of being driven unidirectionally to the boundary, a bulk-localized wave packet becomes confined to oscillate within the bulk. Under a strong field, these oscillations can fully return the particle to its initial position, manifesting as Stark localization.

To illustrate the influence of the electric field in more detail, we consider the evolution of a particle initially localized at the origin with $\psi_{\mathbf{r}^\prime}(0) = \delta_{\mathbf{r}^\prime,0}$. The time-dependent wave function then reduces to
\begin{align}\label{SpeSM}
	\psi_{x,y}(t) = & \exp(g_x x + g_y y) (-i)^{x+y} \mathcal{J}_x(z_x(t)) \mathcal{J}_y(z_y(t)) \notag \\
	& ~~\times \exp[-i(x F_x t/2 + y F_y t/2 )],
\end{align}
with the corresponding probability density
\begin{align}\label{Spe3SM}
	P_{x,y}(t) =  \exp[2(g_x x + g_y y)] \mathcal{J}^2_x(z_x(t)) \mathcal{J}^2_y(z_y(t))/ \mathcal{N}.
\end{align}
The center-of-mass motion is given by
\begin{align}\label{Spe4SM}
	\langle \mathbf{r} (t) \rangle = \sum_\mathbf{r}  \mathbf{r} \, P_{\mathbf{r}}(t).
\end{align}

\begin{figure*}[!tb]
	\centering
	\includegraphics[width=17.8cm]{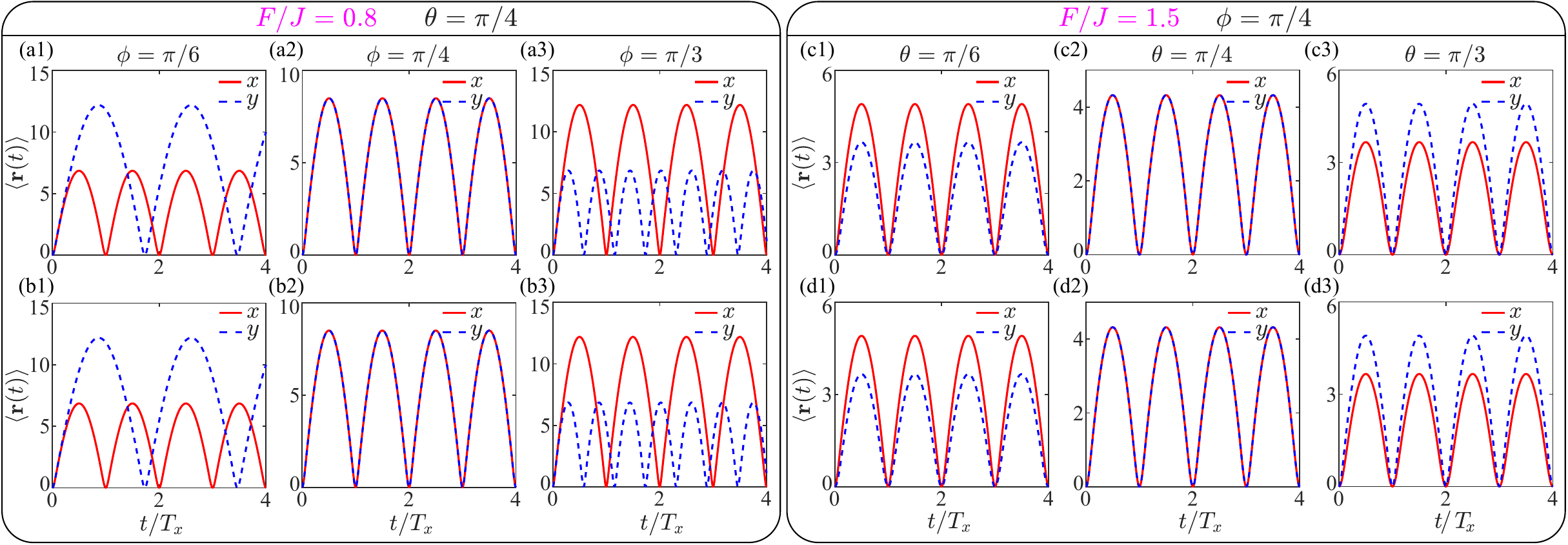}
	\caption{Time-resolved center-of-mass   $\langle \mathbf{r}(t) \rangle$ obtained numerically (top panels), and analytically (bottom panels), for an initial state $\psi_{\mathbf{r}^\prime}(0) = \delta_{\mathbf{r}^\prime,0}$ centered at the origin on $61\times61$ square lattice. Left: $F/J=0.8$ with $\theta=\pi/4$ for different values of $\phi$. Right: $F/J=1.5$ with $\phi=\pi/4$ for different values of $\theta$. Time is measured in unit of $T_x = 2\pi/F_x$. Other parameter used is $g/J=1.0$.}\label{FigS1}
\end{figure*}

Figure \ref{FigS1} compares direct numerical simulations (top panels) with exact analytical results (bottom panels), showing excellent agreement across all parameter regimes. The wave packet does not drift toward the boundary but instead exhibits modified Bloch oscillations induced by the nonreciprocal hopping. Even in the regime of strong nonreciprocity and weak electric field, the Stark effect produces oscillatory motion, causing the wave packet to periodically return to its initial position for $\phi = \pm \pi/4$, or to positions close to the initial one for $\phi \neq \pm \pi/4$. This behavior demonstrates that the electric field alone suppresses the NHSE.

\section{Expansion of the full Hamiltonian in the biorthogonal Wannier-Stark states}\label{AppendixB}

In this section, we analytically construct the biorthogonal Wannier–Stark states of the disorder-free system in the thermodynamic limit and employ them to expand the full non-Hermitian Hamiltonian. This representation provides a transparent picture of how disorder couples localized Stark states in the presence of nonreciprocal hopping. 

We begin with the Hermitian Hamiltonian $\hat{H}^\prime$ in Eq.~(\ref{HamilTSM}). Expanding its eigenstate as
\begin{align}\label{PhiM}
	\ket{\Phi} = \sum_{\mathbf{r}} \phi_{\mathbf{r}} \ket{\mathbf{r}},
\end{align}
the eigenvalue equation $\hat{H}^\prime \ket{\Phi} = E \ket{\Phi}$ gives
\begin{align}\label{SchEqM}
	E \phi_{\mathbf{r}} = J \sum_{\boldsymbol{\delta} = \hat{\mathbf{x}},\hat{\mathbf{y}}} [\phi_{\mathbf{r}+\boldsymbol{\delta}} + \phi_{\mathbf{r}-\boldsymbol{\delta}}] + (\mathbf{F}\cdot\mathbf{r}) \phi_{\mathbf{r}}.
\end{align}

Using the separable ansatz $\phi_{x,y} = \phi(x) \chi(y)$, the problem is reduced to a 1D problem along either $x$ or $y$ directions, whose solutions are the standard Wannier–Stark states,
\begin{align}\label{SolM}
	\phi_n(x) = \mathcal{J}_{x-n}(\gamma_x),\quad \chi_n(y) = \mathcal{J}_{y-n}(\gamma_y),
\end{align}
with eigenenergies
\begin{align}\label{EM}
	E_\alpha^{(n)} = F_\alpha n,\quad n\in\mathbb{Z},
\end{align}
where $\gamma_\alpha = -2J/F_\alpha$ for $\alpha = x,y$. 

Combining both $x$ and $y$ directions, the 2D Wannier-Stark states of $\hat{H}^\prime$, in the absence of disorder, are given by
\begin{align}\label{WSstateM}
	\ket{\Phi_{m,n}} = \sum_{x,y} \mathcal{J}_{x-m}(\gamma_x) \mathcal{J}_{y-n}(\gamma_y) \ket{x,y},
\end{align}
with the ladder energy (i.e., eigenenergy in the thermodynamic limit) being
\begin{align}\label{LadderM}
	E_{m,n}=F_x m+F_y n, \qquad m,n\in\mathbb{Z}.
\end{align}

The Bessel function $\mathcal{J}_{\alpha-n}(|\gamma_\alpha|)$ dictates the localization, being primarily confined to the interval $|\alpha-n| < |\gamma_\alpha|$ and decaying rapidly outside it, with the asymptotic form $\mathcal{J}_{\alpha-n}(|\gamma_\alpha|) \sim |\gamma_\alpha|^{\abs{\alpha-n}}$ for $\abs{\alpha-n} \gg |\gamma_\alpha|$. Consequently, the   state  in Eq.~(\ref{WSstateM}) is exponentially localized along the $x$ and $y$ directions, with a localization length of $2J/F_\alpha$ along the corresponding direction.

Transforming back to the original non-Hermitian frame, the right and left Wannier–Stark eigenstates acquire the exponential gauge factor from Eq.~(\ref{SSM}) with
\begin{align}\label{WannierStateNHRSM}
	\ket{\Psi_{m,n}} = \sum_{x,y} \exp(g_x x + g_y y) \mathcal{J}_{x-m}(\gamma_x) \mathcal{J}_{y-n}(\gamma_y) \ket{x,y},
\end{align}
\begin{align}\label{WannierStateNHLSM}
	\ket{\bar{\Psi}_{m,n}} = \sum_{x,y} \exp(-g_x x - g_y y) \mathcal{J}_{x-m}(\gamma_x) \mathcal{J}_{y-n}(\gamma_y) \ket{x,y}.
\end{align}

Using the Bessel identity $\sum_x \mathcal{J}_{x-m}(z)\mathcal{J}_{x-m'}(z) = \delta_{m,m'}$, one verifies the biorthonormality relation 
\begin{align}
	\langle\bar{\Psi}_{m,n}|\Psi_{m',n'}\rangle = \delta_{m,m'}\delta_{n,n'}.
\end{align}
Completeness is established by multiplying the Hermitian completeness relation $\sum_{m,n} \ket{\Phi_{m,n}} \bra{\Phi_{m,n}} = \hat{I}$ on the left by $\hat{S}^{-1}$ and right by $\hat{S}$, giving
\begin{align}\label{CompletenessSM}
	\sum_{m,n} \ket{\Psi_{m,n}} \bra{\bar{\Psi}_{m,n}} = \hat{I}.
\end{align}

In the biorthogonal Wannier--Stark basis, the disorder-free Hamiltonian is diagonal,
\begin{align}\label{H0SM}
	\hat{H}_\textrm{f} = \sum_{m,n} E_{m,n} \ket{\Psi_{m,n}} \bra{\bar{\Psi}_{m,n}},\quad E_{m,n} = F_x m + F_y n.
\end{align}

We now include onsite disorder, $\hat{V}=\sum_{\mathbf{r}} \xi V_\mathbf{r}  \hat{c}_\mathbf{r}^\dagger \hat{c}_\mathbf{r}$, whose matrix elements in the biorthogonal Wannier--Stark basis are
\begin{align}
	V_{(m,n),(m',n')} = \xi \sum_{x,y} V_{x,y}& \mathcal{J}_{x-m}(\gamma_x) \mathcal{J}_{x-m'}(\gamma_x) \notag \\
	& \times \mathcal{J}_{y-n}(\gamma_y) \mathcal{J}_{y-n'}(\gamma_y).
\end{align}
The full Hamiltonian can therefore be written as
\begin{align}\label{fullHSM}
	\hat{H} = & \sum_{m,n} E_{m,n} \ket{\Psi_{m,n}} \bra{\bar{\Psi}_{m,n}} \notag \\
	& + \sum_{m,n} \sum_{m^\prime,n^\prime} V_{(m,n),(m^\prime,n^\prime)} \ket{\Psi_{m,n}} \bra{\bar{\Psi}_{m^\prime,n^\prime}}.
\end{align}

In this representation, the clean system forms decoupled Stark ladders, while onsite disorder induces effective coupling between exponentially localized Wannier–Stark states.

\section{Ultra-long-time dynamics}\label{AppendixC}
In the main text, we have shown that the interplay among the NHSE, disorder, and an external electric field enables control over the localization direction of wave packets, allowing arbitrary positioning of boundary localization for evolution times up to $t_\textrm{f} = 20 T_B$ by tuning the directions of the electric field and the nonreciprocal hopping. To further clarify the localization behavior, we extend the dynamical evolution into the ultra-long-time regime up to $t_\textrm{f} = 2000 T_B$.

\begin{figure*}[!tb]
	\centering
	\includegraphics[width=17cm]{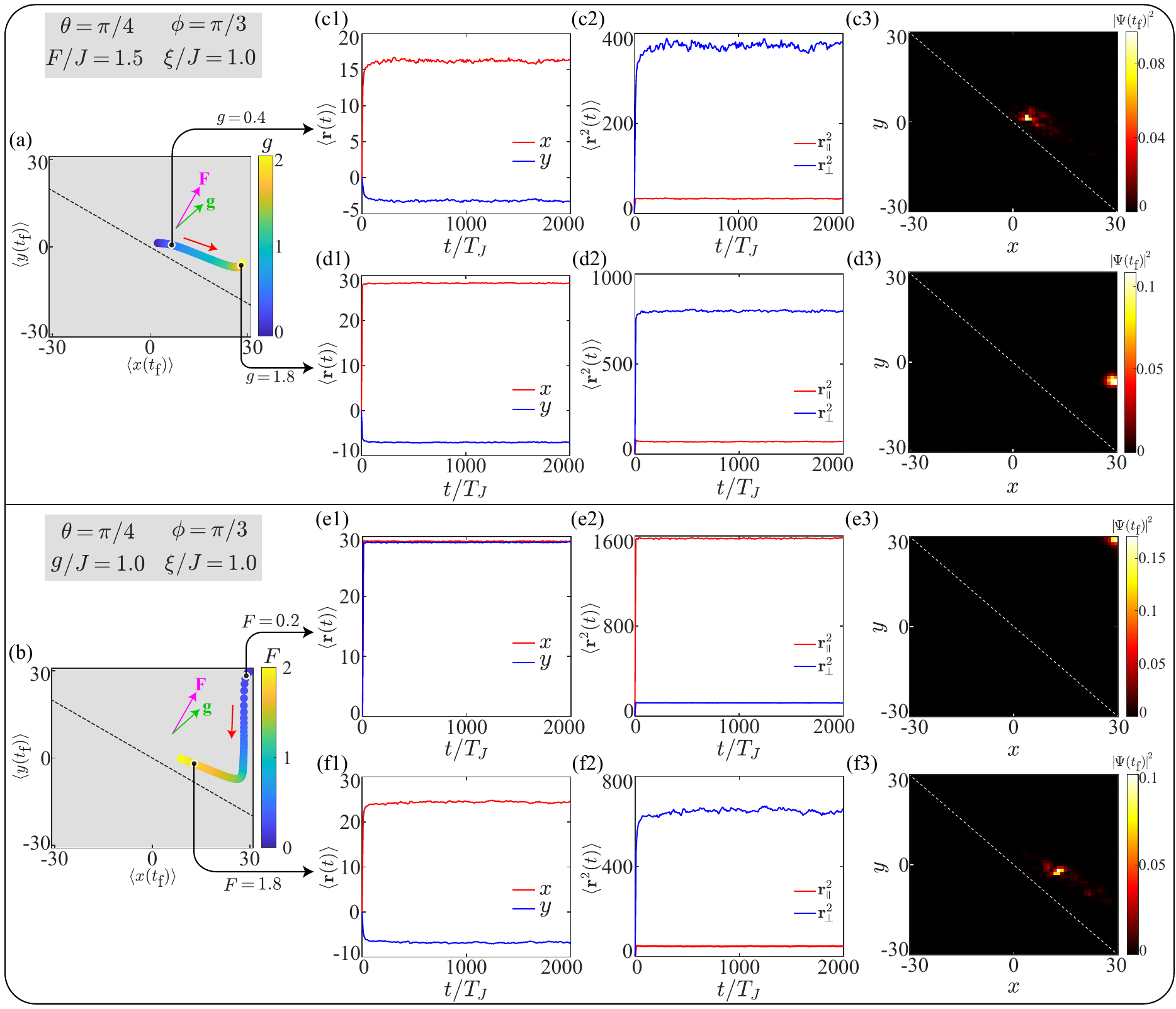}
	\caption{Trajectories of the center of mass, $\langle x(t_\textrm{f}) \rangle$ and $\langle y(t_\textrm{f}) \rangle$, of an origin-centered Gaussian wave packet after long-time evolution under nonreciprocal hopping, electric field, and random on-site disorder, as a function of nonreciprocity strength $g$ (a), and field strength $F$ (b). Panels (c1–f3) display the time-resolved center-of-mass motion $\langle \mathbf{r}(t) \rangle$, the second moment $\langle \mathbf{r}^2(t) \rangle$, and the corresponding probability density distributions after ultra-long-time evolution for the chosen parameter sets indicated in (a) and (b). Except for the varied parameters, others are fixed at $g/J=1.0$, $F/J=1.5$, $\xi/J=1.0$ with $\theta=\pi/4$ and $\phi = \pi/3$. All results are averaged over 100 disorder realizations. }\label{FigS4}
\end{figure*}

Figures~\ref{FigS4}(a) and \ref{FigS4}(b) show the long-time center-of-mass displacement $\langle \mathbf{r}(t_\mathrm{f}) \rangle$ of an initially localized Gaussian wave packet as a function of the nonreciprocity strength $g$ and the field strength $F$, respectively. From each parameter sweep, we select two representative cases, one leading to bulk localization and the other to boundary localization, highlighted by the black arrows in Figs.~\ref{FigS4}(a,b).

The corresponding time-resolved dynamics, shown in Figs.~\ref{FigS4}(c1–f3), reveal that the wave packet first undergoes a drift stage before settling into a localized configuration with only weak residual oscillations. The nearly constant values of both $\langle \mathbf{r}(t) \rangle$ and $\langle \mathbf{r}^2(t) \rangle$ indicate that spreading is strongly suppressed, while the final probability density profiles confirm that localization occurs either in the bulk or at the boundary depending on the chosen parameters. These results demonstrate that our control over mode localization remains robust under ultra-long-time evolution.

\section{Liouvillian Dynamics}\label{AppendixD}

Although the main text focuses on dynamics generated by non-Hermitian Hamiltonians, analogous effective dynamics arise naturally in open quantum systems. Within the Born-Markov approximation \cite{Daley2014, Sieberer2016}, the density matrix $\hat{\rho}$ evolves according to the Lindblad master equation \cite{Scully1997, Breuer2007, Agarwal2012}
\begin{align}\label{LindbladSM}
	\frac{d \hat{\rho}}{d t} = \mathcal{L} \hat{\rho} = -i [\hat{H}_0, \hat{\rho}] + \sum_{\mu} \left( \hat{L}_\mu \hat{\rho} \hat{L}^\dagger_\mu - \frac{1}{2} \{\hat{L}^\dagger_\mu \hat{L}_\mu, \hat{\rho}\}\right),
\end{align}
where $\hat{H}_0$ denotes the tight-binding Hamiltonian for the Hermitian system, and the Lindblad operators $\hat{L}_\mu$ encode environment-induced quantum jumps.

To characterize the system’s dynamics, it is convenient to monitor the single-particle correlation function of a $d$-dimensional dissipative lattice, $\Delta_{\mathbf{r},\mathbf{r}^\prime}(t) = \Tr [\hat{c}^\dagger_\mathbf{r} \hat{c}_{\mathbf{r}^\prime} \hat{\rho}(t)]$, whose time evolution is given by
\begin{align}\label{DeltaEvoSM}
	\frac{d \Delta_{\mathbf{r},\mathbf{r}^\prime}(t)}{d t} = \Tr \left[\hat{c}^\dagger_\mathbf{r} \hat{c}_{\mathbf{r}^\prime} \frac{d \hat{\rho}}{dt}\right].
\end{align}
Substituting the Eq.~(\ref{LindbladSM}) into Eq.~(\ref{DeltaEvoSM})  yields
\begin{align}\label{MeanSM}
	&\frac{d \Delta_{\mathbf{r},\mathbf{r}^\prime}(t)}{dt} = -i \Tr \left([\hat{c}^\dagger_\mathbf{r} \hat{c}_{\mathbf{r}^\prime}, \hat{H}_0] \hat{\rho}(t )\right) \notag \\
	& ~~+ \sum_{\mu} \Tr \left[\left(\hat{L}^\dagger_\mu [\hat{c}^\dagger_\mathbf{r} \hat{c}_{\mathbf{r}^\prime}, \hat{L}_\mu] + \frac{1}{2} [\hat{L}^\dagger_\mu \hat{L}_\mu, \hat{c}^\dagger_\mathbf{r} \hat{c}_{\mathbf{r}^\prime}]\right) \hat{\rho}(t) \right].
\end{align}
Here, the Hermitian single-particle Hamiltonian is written as 
\begin{align}\label{MeanFinalSM2}
	\hat{H}_0 = \sum_{\mathbf{j},\mathbf{k}} h_{\mathbf{j}\mathbf{k}} \hat{c}^\dagger_\mathbf{j} \hat{c}_{\mathbf{k}},
\end{align}
and the Lindblad operators incorporate both gain and loss processes. Each jump operator is either a gain channel, $\hat{L}_\mu = \hat{L}^g_\mu$, or a loss channel, $\hat{L}_\mu = \hat{L}^l_\mu$, with
\begin{align}\label{MeanFinalSM3}
	\hat{L}^g_\mu = \sum_{\mathbf{k}} D_{\mu \mathbf{k}}^g \hat{c}^\dagger_\mathbf{k},~~~\hat{L}^l_\mu = \sum_{\mathbf{k}} D_{\mu \mathbf{k}}^l \hat{c}_\mathbf{k}.
\end{align}

Defining $M^g_{\mathbf{j} \mathbf{k}} = \sum_{\mu} D^{g\ast}_{\mu \mathbf{j}} D^g_{\mu \mathbf{k}}$ and $M^l_{\mathbf{j}\mathbf{k}} = \sum_{\mu} D^{l\ast}_{\mu \mathbf{j}} D^l_{\mu \mathbf{k}}$, one obtains the compact evolution equation
\begin{align}\label{MeanFinalSM}
	\frac{d \Delta_{\mathbf{r},\mathbf{r}^\prime}(t)}{dt} = \left[M^g + X \Delta(t) + \Delta(t) X^\dagger \right]_{\mathbf{r},\mathbf{r}^\prime},
\end{align}
with 
\begin{align}\label{MeanFinal2SM}
	X = i h^T - \frac{1}{2} [M^g + (M^l)^T].
\end{align}

We now specify a two-dimensional lattice governed by the Hermitian Hamiltonian
\begin{align}\label{HerHSM}
	\hat{H}_0 =  & \sum_{x,y} \left(J \cosh g_x \hat{c}^\dagger_{x+1,y} \hat{c}_{x,y} + J \cosh g_y \hat{c}^\dagger_{x,y+1} \hat{c}_{x,y} + \textrm{H.c.}\right) \nonumber \\
	&+\sum_{x,y} \left[(F_x x + F_y y) \hat{c}^\dagger_{x,y} \hat{c}_{x,y} + \xi   V_{x,y} \hat{c}^\dagger_{x,y} \hat{c}_{x,y} \right],
\end{align}
where $\hat{c}_{x,y}$ ($\hat{c}^\dagger_{x,y}$) annihilates (creates) a particle at site ($x,y$). The coefficients $J \cosh g_x$ and $J \cosh g_y$ specify the hopping amplitudes along the $x$ and $y$ directions, with $g_x = g \cos \theta$ and $g_y = g \sin \theta$, $F_x$ and $F_y$ are static electric field amplitudes along the $x$ and $y$ directions, and $\xi V_\mathbf{r}$ represents the random on-site potential with  $V_\mathbf{r} \in [-1/2,1/2]$ and disorder strength $\xi$.  

We introduce two channels of nonlocal  dissipative processes acting on nearest neighbors along $x$ and $y$ directions, described by the Lindblad operators
\begin{align}\label{jumps1SM}
	\hat{L}_{x,y}^1 = \sqrt{2J\sinh g_x} (e^{g_x/2} \hat{c}_{x+1,y} + i e^{-g_x/2} \hat{c}_{x,y}),
\end{align}
\begin{align}\label{jumps2SM}
	\hat{L}_{x,y}^2 = \sqrt{2J\sinh g_y} (e^{g_y/2} \hat{c}_{x,y+1} + i e^{-g_y/2} \hat{c}_{x,y}).
\end{align}

Using Eqs.~(\ref{MeanFinalSM}) and (\ref{MeanFinal2SM}), one finds that the correlation matrix obeys the closed equation
\begin{align}\label{DeltaSM}
	\frac{d \Delta}{d t} = i (H\Delta -\Delta H^\dagger),
\end{align}
where $H = h^T + \frac{i}{2} (M^l)^T$ takes the explicit form
\begin{align}\label{finalResSM}
	&H  = \sum_{x,y} \left(J e^{g_x} \ket{x+1,y} \bra{x,y} + J e^{-g_x} \ket{x,y} \bra{x+1,y}\right) \nonumber \\
	& ~ + \sum_{x,y} \left(J e^{g_y} \ket{x,y+1} \bra{x,y} + J e^{-g_y} \ket{x,y} \bra{x,y+1}\right) \nonumber \\
	& ~ + \sum_{x,y} \left[(F_x x + F_y y) \ket{x,y} \bra{x,y} + \xi   V_{x,y} \ket{x,y} \bra{x,y} \right] \nonumber \\
	& ~ + \sum_{x,y} \left[iJ(\sinh 2g_x + \sinh 2g_y)\ket{x,y} \bra{x,y}\right].
\end{align}
This Hamiltonian describes a two-dimensional extension of the Hatano–Nelson model subjected to a static electric field and random on-site disorder. It coincides with the Hamiltonian studied in the main text, apart from an inessential homogeneous damping term appearing in the final term.

Equation (\ref{DeltaSM}) is solved directly as
\begin{align}\label{finalRes2SM}
	\Delta(t) = e^{iHt} \Delta(0) e^{-iH^\dagger t}.
\end{align}
For a particle initially localized at the origin, $\Delta(0) = \ket{\mathbf{0}} \bra{\mathbf{0}}$, the correlation at time $t$ becomes $\Delta(t) = \ket{\psi(t)} \bra{\psi(t)}$ with $\ket{\psi(t)} = e^{iHt} \ket{\mathbf{0}}$. The normalized probability distribution of surviving particles is therefore
\begin{align}\label{ProbabilitySM}
	P_\mathbf{r}(t) = \Delta_{\mathbf{r},\mathbf{r}}(t) / \sum_{\mathbf{r}} \Delta_{\mathbf{r},\mathbf{r}}(t) = \abs{\bra{\mathbf{r}} \ket{\psi(t)}}^2 / \bra{\psi(t)} \ket{\psi(t)}.
\end{align}

This matches precisely the evolution generated by the effective non-Hermitian Hamiltonian $H$, demonstrating that the open quantum system reproduces the same dynamical behavior as the non-Hermitian lattice model studied in the main text, including the effects of the electric field and disorder.


%

\end{document}